\documentclass[12pt]{iopart}

\usepackage{graphicx}
\usepackage{iopams}  

\begin{document}




\title{Nuclear and detector sensitivities\\for neutrinoless double beta-decay experiments}


\author{Hiroyasu Ejiri }

\address{ Research Center for Nuclear Physics, Osaka University, Osaka 567-0047, Japan\\
E-mail ejiri@rcnp.osaka-u.ac.jp
}
\begin{abstract}
Neutrinoless double beta-decay (DBD) is of current interest in high-sensitivity frontiers of particle physics. The decay is very sensitive to Majorana neutrino masses,  neutrino CP phases,  right-handed weak interactions and others, which are beyond the standard electro-weak model. DBDs are actually ultra-rare events, and thus DBD  experiments with ultra-high sensitivity are required. Critical discussions are presented on nuclear  and  detector sensitivities for high-sensitivity DBD experiments to study the neutrino masses in the normal and inverted mass-hierarchies. \\


~~~~~~~~~~~neutrino mass, Majorana neutrino, nuclear matrix element. 
\end{abstract}




\section{Introduction}
Neutrinoless double beta-decay (DBD) is of current interest in sensitivity frontiers of particle physics. The decay violates the lepton number conservation law, and thus DBD is beyond the standard electro-weak model.  DBD is very sensitive to  the Majorana nature (neutrino=anti-neutrino) of neutrino, the absolute neutrino-mass scale and  the neutrino-mass hierarchy,  the neutrino CP-phases, the possible right-handed weak interactions and  others, which are beyond the standard model. Accordingly, it has been used to study these fundamental questions on the neutrino and  the weak interaction for decades.
Recent experimental and theoretical DBD works  are discussed in the review articles and references therein \cite{eji05,avi08,eji10,ver12,ver16}.

In fact, neutrinoless DBDs are due to various modes such as the light neutrino-mass mode, the right-handed weak current mode, the SUSY particle exchange mode, and the others, which are all beyond the standard model \cite{eji05,ver12} . In the present work, we discuss mainly the neutrinoless DBD due to the light neutrino-mass mode, which is of current interest. 

DBD, however, is a very  low-energy and ultra-rare decay. The energy is of the order of $E_{\beta \beta }\approx$ 10$^{-3}$ GeV and the decay rate is in the range of  $T^{0\nu}_r\approx $ 10$^{-27}$ - 10$^{-30} $ per year (y), depending on the neutrino-mass and the nuclear matrix element(NME). 
The neutrino-mass to be studied is around $m$=3 meV and 30 meV in cases of the normal mass-hierarchy (NH) and the inverted mass-hierarchy, respectively.  Then,  low-background high-sensitivity DBD detectors are required to search for such low-energy and ultra-rare DBD signals \cite{eji05,ver12}. 

Many experimental groups are working hard for high sensitivity DBD experiments to search for the small Majorana neutrino-mass and the others beyond the standard model. Current lower limits on the DBD half-lives are around  10$^{26}$ y for both $^{76}$Ge and $^{136}$Xe .
The neutrino mass limits are, respectively, 600 meV/$M(76)$ and 230 meV/$M(136)$ with $M(76)$ and $M(136)$ being the nuclear matrix elements (NMEs) for $^{76}$Ge and $^{136}$Xe, respectively. So they are in the range of 300-100 meV if NMEs are around 2. See section 4 for  discussions on DBD experiments.
Then one needs to improve the  sensitivity by factors 5 and 50 to access the IH and NH mass regions, respectively. 
In fact, the DBD sensitivity to the neutrino mass depends much on NME. 

 \begin{figure}[htb]
\begin{center} 
\hspace{-0cm}
\includegraphics[width=0.6 \textwidth]{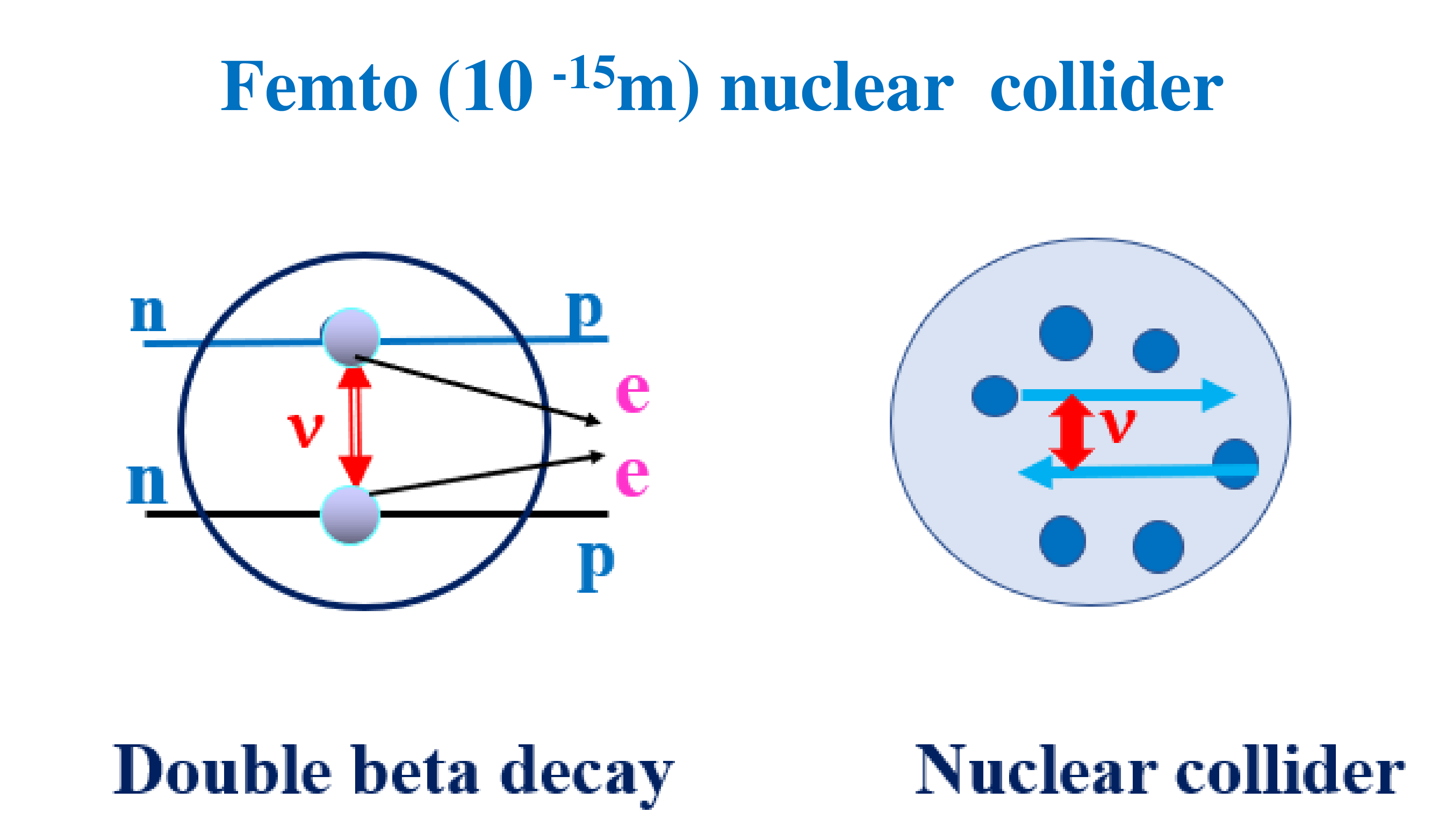}
\end{center} 
\caption{Schematic views of  neutrinoless double $\beta$-decay with  Majorana neutrino exchange (left hand  panel) and nuclear collider (right handed panel).
}
\label{fig:figure1}  
\end{figure}

DBD with a light neutrino-mass mode is considered as a neutrino-exchange process, where a light Majorana neutrino is exchanged between two neutrons in the DBD nucleus. Here the nucleus acts as a  high-luminosity micro-collider of  two neutrons, as shown in Fig.1. The neutrino-exchange cross-section for the  IH neutrino is as small as $\sigma \approx $ 10$^{-83}$ cm$^2$ because of the second-order weak process and the small neutrino mass \cite{ver12,eji00,eji19}. The luminosity of one nucleus is as high as $L\approx$10$^{48}$ cm$^{-2}$s$^{-1}$ because of the very small ( a few fm) distance between the two neutrons in the nucleus. The summed luminosity for 1 ton DBD nuclei is $L\approx $10$^{76}$cm$^{-2}$s$^{-1}$. Thus one gets 2-3 DBD signals per year for the IH neutrino exchange by using a large DBD detector with 1 ton  DBD isotopes. 

The DBD neutrino-mass sensitivity to search for the small neutrino  mass is defined as the minimum neutrino mass $m_m$ to be measured by the DBD experiment. It is given by a product of a nuclear sensitivity and a detector sensitivity. The nuclear sensitivity depends on nuclear parameters such as the nuclear phase space and the NME, while the detector sensitivity depends on the detector parameters such as the total number of the DBD isotopes, the enrichment, the exposure time, the detection efficiency and the backgrounds. The minimum neutrino mass to be measured and the nuclear and detector sensitivities are discussed in the review articles \cite{eji05,ver12,eji19}.

The present report aims at critical discussions on the nuclear  and detector sensitivities to search for the
 ultra-rare DBD events associated with very small  IH and NH masses.  Critical discussions on DBD NMEs is presented  elsewhere \cite{eji20}.

\section{Neutrino mass sensitivity for DBD experiment}

The neutrinoless DBD transition rate is expressed in terms of the axial-vector weak coupling $g_A$=1.27 in units of the vector coupling of $g_V$ as \cite{eji05,ver12}
\begin{equation}
T_r^{0\nu}=g_A^{4}G^{0\nu}|M^{0\nu}f(\nu)|^2,
\end{equation}
where $G^{0\nu}$ is the  phase space volume,  $M^{0\nu}$ is the  NME  and $f(\nu)$ stands for the effective neutrino mass and the other neutrino and weak interaction parameters beyond the standard model.  These three quantities are associated with the kinematic factor, the nuclear physics factor and the particle physics factor, respectively. 

 In case of the light  neutrino-mass process, $f(\nu)$ is given by the effective neutrino mass of $m=\sum |U_i|^2 m_ie^{\alpha _i}$ with $m_i$ $U_i$ and $\alpha _i$ being the ith neutrino mass, the mixing coefficient  and the phase, respectively. In case of the right-handed weak-boson process  $f(\nu)$ includes the term  $k(M_W^L/M_W^R)^2 $ with $M_W^L$ and $M_W^R$ being the let-handed and right-handed weak-boson masses, respectively \cite{eji05,ver12}. 

The DBD NME for the light neutrino-mass mode is expressed in terms of the Gamow-Teller (GT), the Fermi (F) and the tensor (T) NMEs as \cite{eji05,ver12,eji19}
\begin{equation}
M^{0\nu}=(\frac{g_A^{eff}} {g_A})^2 [M(GT)+(\frac{g_V^{eff}}{g_A^{eff}})^2M(F)+M(T)],
\end{equation}
where $g_A^{eff}$ and $g_V^{eff}$ are, respectively,  the effective axial-vector and vector  couplings, and  $g_A$ and $g_V$ are the axial-vector and vector couplings for a free nucleon. $M(GT)$, $M(F)$ and $M(T)$ are, respectively,  the model NMEs for Gamow-Teller, Fermi, and tensor transition operators.  

The GT and T NMEs are very sensitive to  nucleonic and non-nucleonic correlations and nuclear medium effects, and thus the ratios $g_A^{eff}/g_A$ and $g_V^{eff}/g_V$ stand for the re-normalization coefficients due to such correlations and nuclear medium effects that are not explicitly included in the nuclear model. Thus they depends on the models and the transition operators. Since the axial-vector NMEs are strongly modified by the strong isobar (quark spin-isospin excitation) and other non-nucleonic correlations, we usually consider explicitly the re-normalization coefficient $g_A^{eff}$, but the re-normalization coefficient $g_V^{eff}$ for the vector coupling should be well considered unless the models include
all correlations and the medium effect \cite{eji00,eji19,eji20}.
The DBD NMEs are discussed in review articles on neutrino nuclear responses \cite{eji00,eji19,eji20}. Theoretical works on DBD NMEs are discussed in reviews \cite{suh98,suh12,eng17}. 

The DBD rate $T_r^{0\nu}$ per year (y) and per ton (t) of the DBD isotope-mass is expressed by using  a neutrino mass-unit $m_0$ as \cite{eji05,ver12,eji19,eji20}
\begin{equation}
T_r^{0\nu}=(\frac{m}{m_0})^2, ~~~ m_0=\frac{m'_0}{M^{0\nu}}
\end{equation}
\begin{equation}
m'_0=7.8 {\rm meV} \frac{ A^{1/2}} {g_A^2 (G^{0\nu})^{1/2}},
\end{equation}
where $A$ is the mass-number of the DBD nucleus and $G^{0\nu}$ is the phase space factor in unit of 10$^{-14}$/y, and   $m_0$ is the neutrino mass that gives the DBD rate $T^{0\nu}_r$=1 per t y. We define $m_0$ as the nuclear sensitivity of the DBD nucleus, being inversely proportional to the ($G^{0\nu})^{1/2} M^{0\nu}$.  Here, high sensitivity to the neutrino mass means high capability of measuring a small neutrino-mass. 
$m'_0$ is the neutrino mass $m_0$ in case of $M^{0\nu}$=1.

The values $m'_0$ for typical DBD nuclei of $^{82}$Se, $^{100}$Mo, $^{116}$Cd and $^{130}$Te, which are of current interest, are all around 40 meV, which are close to the upper bound of the IH neutrino mass. The values for $^{76}$Ge and $^{150}$Nd are, respectively, larger and smaller  by a factor 2 than the value of 40 meV because of the smaller and larger phase space volumes.  



In order to study the neutrino mass, the number of the DBD signals  is required to be larger than the fluctuation $\delta$ of the number of the background signals.  Then it is expressed as \cite{eji05,eji19,eji20}
\begin{equation}
T_r^{0\nu} \eta \epsilon NT \ge \delta, ~~~~~ \delta = \delta_0 \times (BNT)^{1/2},
\end{equation}
where  $N$ is the total DBD isotope-mass in units of t, $T$ is the exposure in units of y, $\delta _0$ is around 2 and $B$ is the number of the backgrounds per  $N$=1 ton per $T$=1 year, $\eta$ is the enrichment coefficient of the DBD isotope and $\epsilon $ is the DBD signal detection-efficiency. Note that the actual DBD isotope-mass is $\eta N$ ton,  the signal yield is $T_r^{0\nu} \eta \epsilon NT$,  and the BG yield is $BNT$. 

Then the minimum effective neutrino-mass to be measured with 90 $\%$ confidence level is expressed in terms of the nuclear sensitivity $m_0$ and the detector sensitivity $d$ as

\begin{equation}
m_m=m_0\times {d},~~~ d=d_0 \times  \eta ^{-1/2} \epsilon^ {-1/2} (NT/B)^{-1/4},
\end{equation}
where $d_0$ is around 1.4.
The neutrino mass to be measured is $m_m$=2$m'_0/M^{0\nu}$ in a typical case of the DBD detector exposure of $NT$= 1 t y, the BG rate of $B$=1/t y, $\epsilon$=0.55, and $\eta$=0.9.   

On the basis of the simple expression of the mass sensitivity $m_m$ as given in eq. (6), key points for high sensitivity DBD experiments are as follows.

1. The neutrino mass to be measured is proportional to 1/$M^{0\nu}$ and  ($B/NT)^{1/4}$. Therefore the DBD nucleus with larger $M^{0\nu}$ by a factor 2 is equivalent to the DBD isotope-mass ($N$) larger by  a factor 16.  Then the detector volume gets more than an order of magnitude larger if $M^{0\nu}$ gets smaller by 40 $\%$. 

2. The minimum mass to be measured depends on the ratio of $(B/N)^{1/4}$. Then the neutrino mass sensitivity is improved by a factor 2 by increasing the DBD isotope-mass ($N$) by a factor 16, or by decreasing the BG rate by a factor 16, or by increasing the DBD isotope-mass by a factor 4 and decreasing the BG rate by a factor 4. 

3. The effective neutrino mass depends on the neutrino mixing-phases, and  is in the regions of $m\approx$17-45 meV and $m\approx$1.5-4 meV, respectively, in cases of IH and NH.  Assuming a typical  NME of $M^{0\nu}$=2,  one needs the DBD detectors with the mass sensitivities  around the nuclear sensitivity of $m_0\approx 20 $ meV and one tenth of that, respectively,  in cases of  IH and NH. Then one needs DBD detectors with $N\approx$ 1 t and $B\approx$ 1 /t y and $N\approx$ 100 t and $B\approx$ 0.01 /t y , respectively, to measure the IH and NH neutrino-masses. 

4.  DBD  isotope enrichment is very effective. Enrichment by a factor 3  is equivalent to the  increase of DBD isotope-mass by a factor 10. Accordingly it is effective to use ton-scale  $80-90\%$ enriched DBD isotopes. 

\section{DBD nuclear and detector sensitivities}

In this section we discuss DBD nuclear and detector sensitivities as given in eq. (6) for DBD experiments to access the IH and NH neutrino masses.

\subsection{Nuclear sensitivity}

The nuclear sensitivity $m_0$ is a ratio of the unit mass $m'_0$ and the NME $M^{0\nu}$.  In fact, $M^{0\nu}$ is so sensitive to the details of the nuclear structures  that  accurate evaluation for $M^{0\nu}$ is very hard.  So we discuss in the present work mainly $m'_0$, and assume that $M^{0\nu}$ is in the region of  1-3.  See the recent review on DBD NMEs \cite{eji19} and references therein for detailed  discussions on $M^{0\nu}$.
\begin{figure}[htb]
\begin{center} 
\hspace{-0cm}
\includegraphics[width=0.7\textwidth]{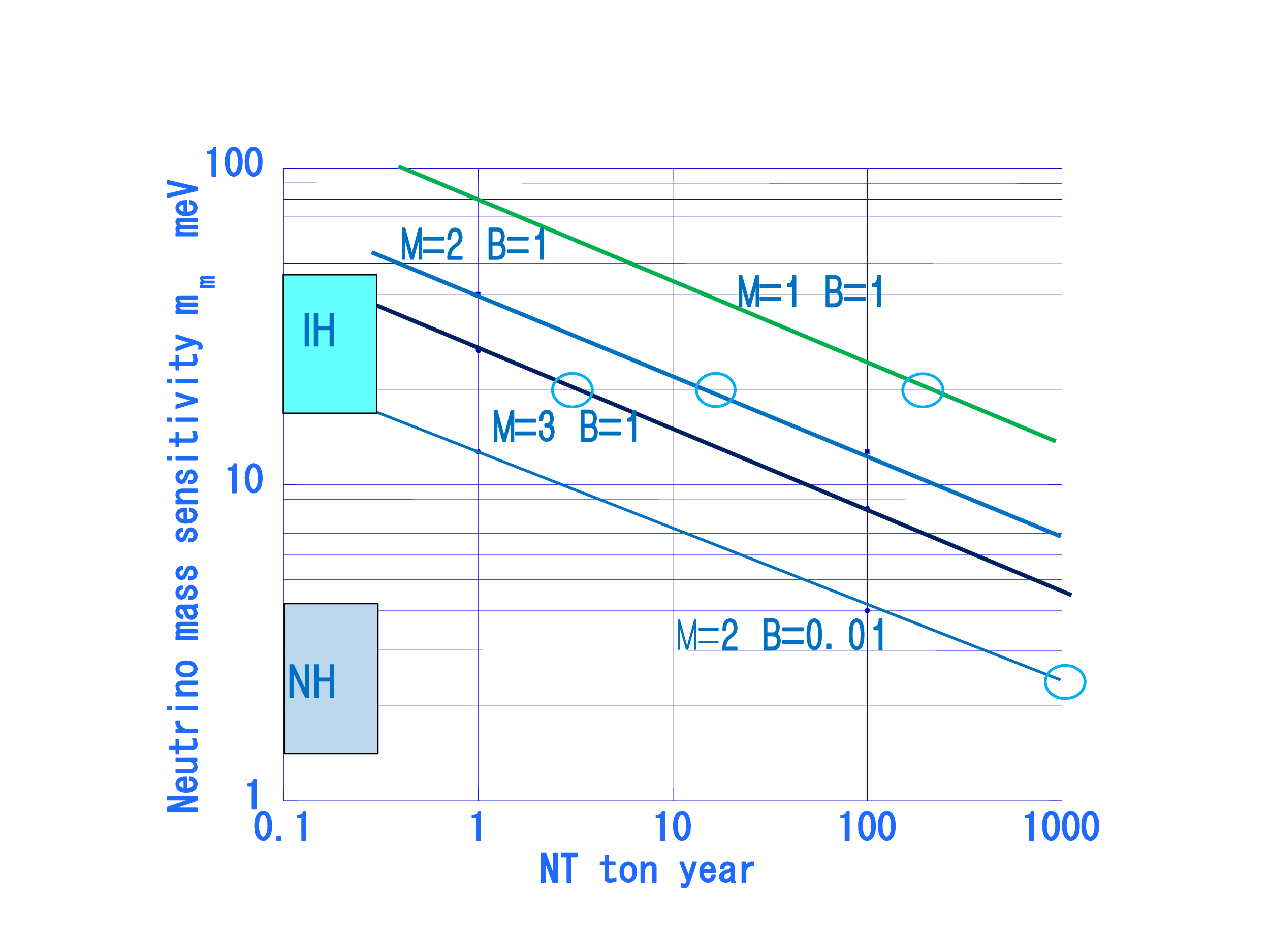}
\end{center} 
\caption{ Neutrino mass sensitivities $m_m$ as a function of the exposure $ NT$.  $M$ and $B$ are the NME and the BG rate per t y.  $d_0$=1.4, $\epsilon$=0.55, $\eta$=0.9. See eq. (6).}.
\end{figure}

The unit mass $m'_0$ is proportional to $ A^{1/2} (G^{0\nu})^{-1/2}$.
The phase space factor $G^{0\nu}$ 
 increases with increase of the DBD $Q$ value. Thus double $\beta ^-$  nuclei with the large $Q$ around 3 MeV are used. The large $Q$ is also effective to reduce much BGs from $^{208}$Tl and $^{214}$Bi, which are two major BGs.  
 The mass-number $A$ dependence reflects the number of the DBD nuclei per ton of the total DBD isotope-mass. 
 The mass sensitivities for a typical DBD nucleus with $m'_0$=40 meV are plotted as a function of the exposure $NT$ in Fig. 2.  
 In order to access the IH mass of 20 meV, one  needs a DBD exposure $NT$ of around 16 t y , i.e.$N\approx$ 5 ton DBD isotopes and $T\approx$3 y measurement  in case of a  typical NME of $M^{0\nu}$=2 and BG rate of $B$=1/t y.  On the other hand, one needs $NT$ of around 3 t y, i.e. $N\approx$1 ton DBD isotopes and $T\approx$3 y measurement in case of $M^{0\nu}$=3, and $NT$ of around 250 t y, i.e.$N\approx$60 ton DBD isotopes and $T\approx$4  y measurement, in case of $M^{0\nu}$=1. So, it is very crucial to know the NME even for designing the DBD detector.


 The mass sensitivities for the enrichment factors of $\eta$=10, 50, 100 $\%$  are plotted as a function of the total DBD isotope-mass $N$ in Fig. 3 

\begin{figure}[htb]
\begin{center} 
\hspace{-0cm}
\includegraphics[width=0.7\textwidth]{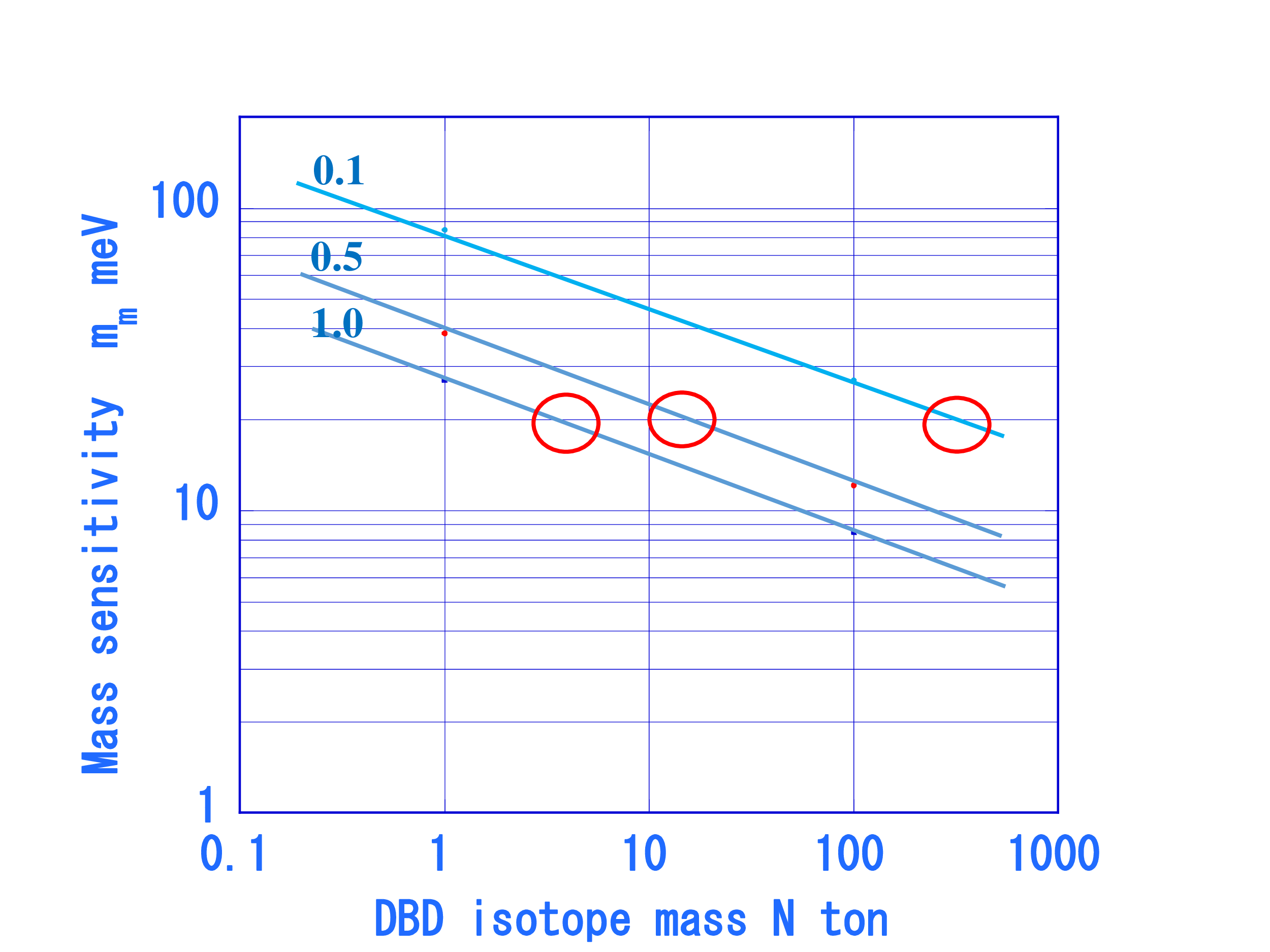}
\end{center} 
\caption{Neutrino mass sensitivities $m_m$ for $\eta$=1.0, 0.5, and 0.1 as a function of the total isotope-mass  $ N$. The exposure time is $T$=5 y, and the nuclear sensitivity of  $m_0$=20 meV, $\epsilon$=0.55, $d_0$=1.4, and $B$=1 /t y. }
\end{figure}

\begin{figure}[htb]
\begin{center} 
\hspace{-0cm}
\includegraphics[width=0.7\textwidth]{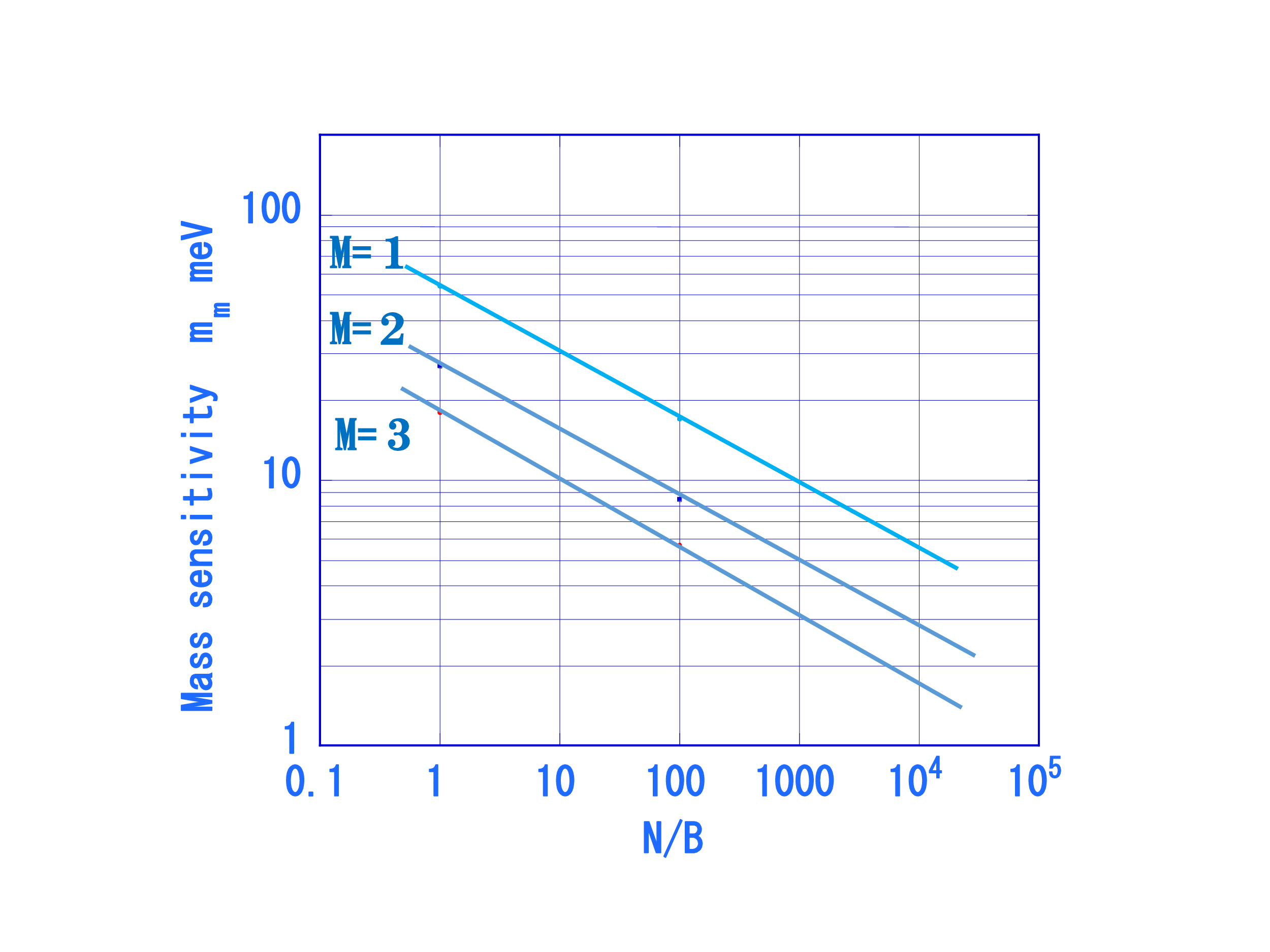}
\end{center} 
\caption{Neutrino mass sensitivities $m_m$ for  $\eta$=0.9, $\epsilon$=0.55, $d_0$=1.4  as a function of the ratio of the total isotope-mass  $ N$ to the BG rate $B$.  The exposure time is $T$=5 y, and the unit mass $m'_0$=40 meV}. 
\end{figure}

In order to access the IH mass of 20 meV, one needs $N$=3.4 t and 4.1 t of the total  DBD isotopes, respectively, in case of $\eta$=100$\%$ and 90$\%$,  while it is $N$=14 t in case of $\eta$=50$\%$ ($\eta N$ = 7 t), and $N$=340 t in case of $\eta$=10$\%$ ($\eta N$ = 34 t). Therefore, enrichment around $\eta \ge 80 \%$ is very effective . The 1 ton DBD total isotope-mass with $\eta \approx 90\%$ is equivalent to the $N\approx$10 ton  of the total DBD isotope-mass with $\eta \approx$ 30$\%$.

\subsection{DBD detector sensitivity and backgrounds}

DBD experiment requires an  ultra-low BG detector since the DBD signal is very low in energy and the event rate  
 is very rare. The neutrino mass-sensitivity for a typical exposure time of $T$=5 y and $\eta$ = 0.9,  $\epsilon$=0.55,
 $d_0$=1.4   is rewritten in therms of the ratio of the DBD isotope-mass $N$ and the BG rate $B$ as 
 \begin{equation}
m_m\approx1.35 \frac{m'_0}{M^{0\nu} (N/B)^{1/4}}.
\end{equation} 

\begin{figure}[htb]
\begin{center} 
\hspace{-0cm}
\includegraphics[width=0.9\textwidth]{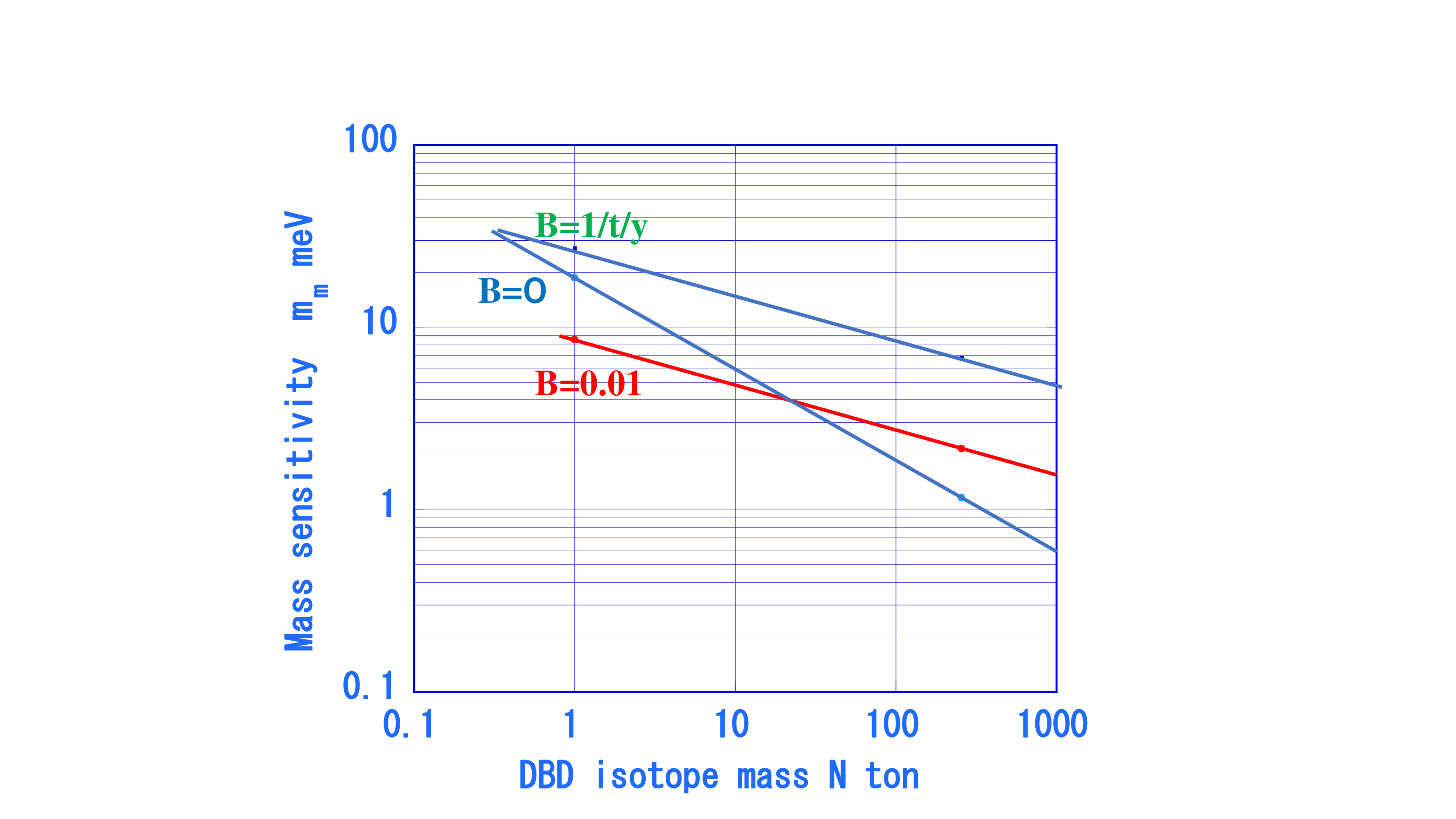}
\end{center} 
\caption{Neutrino mass sensitivities $m_m$ as a function of the DBD isotope-mass  $ N$ in cases of  $B$=1, 0.01, and 0.  The exposure time is $T$=5 y.  The unit mass $m'_0$=40 meV and $M^{0\nu}$=2, $\epsilon$=0.55 and $\eta$=0.9.} 
\end{figure}

The mass sensitivities for the $m'_0$= 40 meV with typical NMEs of $M^{0\nu}$=2 and $M^{0\nu}$=3 are simply given as $ m_m\approx$27  meV/$(N/B)^{1/4}$ and $ m_m\approx$18 meV/$(N/B)^{1/4}$. They are shown as a function of $N/B$ in Fig. 4.  So, in case of $M^{0\nu}\approx$2, DBD detectors
to search for the IH mass of 20-30 meV and the NH mass of 2-3 meV are required to be  large isotope-mass and low-BG detectors  with $N/B\approx$1 and $N/B\approx 10^{4}$, respectively.

Let us see how the mass sensitivity changes if one  gets a  BG free ($B$=0) detector. In this case, the mass sensitivity is derived by requiring the signal yield $\ge$2.3  as \cite{ver12}  
\begin{equation}
m_m=1.5\times m_0  \epsilon^{-1/2} \eta^{-1/2}  (NT)^{-1/2}.
\end{equation}
 The mass sensitivities for three BG cases of $B$=1, 0.01, and 0 are compared in Fig. 5. Then the DBD
  isotope-mass $N$ required to access the IH mass of 20 meV  is 1 t, while  the DBD isotope-mass to access the NH mass of 2 meV is 100 t.  They are a factor 3 smaller than those for the detectors with $B$=1 and 0.01. Thus one needs 1ton-scale  and 100 ton-scale DBD isotope-masses to access the IH and NH neutrino masses even by using the BG-free detectors. 

The DBD signal is expected to appear as a sharp peak at the energy of $E=Q$ in the energy spectrum. The peak width
is around the FWHM (full width half maximum) of the detector. Thus one usually sets the energy window of $\Delta E $= 2 FWHM as the region of interest, ROI.  
 
 BGs in ROI are mostly due to $\beta $-rays and Compton-scattered $\gamma $ rays, which are continuum spectra. 
 Backgrounds due to the solar-neutrino interaction, which get serious in the NH mass search, are also continuum \cite{eji14,eji17}.
Then the BG rate at the energy window of ROI is  proportional to FWHM.  Since the mass sensitivity is proportional to $(B/N)^{1/4}$, improvement of the energy resolution by an order of magnitude is equivalent to increase of the DBD isotope-mass by the same order of magnitude. 

The mass sensitivity depends on the detection efficiency as $m_m$ = $k \epsilon ^{-1/2}$. The efficiency  includes 
all efficiencies associated with the energy and PSA windows/cuts and analyses to select the DBD signals and to reject BG ones. It is around $\epsilon\approx $0.5-0.6.  Severe cuts to reject BGs reduce both the efficiency $\epsilon$ and the BG rate $B$. The decrease of $\epsilon$ by 20$\%$ is compensated if  $B$ gets smaller by 40 $\%$.

\section{Remarks and discussions on DBD detectors}
There are many  high-sensitivity DBD experiments and future plans to access the IH mass  around 20 meV and  the NH mass around 2 meV.  The detailed reports on the present status on individual experiments and plans are given in their reports and  detailed discussions on them  are given in the reviews \cite{eji05,ver12,ver16,eji19} and references therein, and also in the recent 2020 conference reports. The present and future DBD experiments are discussed there, see for example ref. \cite{det20}. So, we  give in this section a few comments on some (not all) possible DBD experiments from view points of the DBD mass sensitivity, i.e. the nuclear and detector sensitivities as given in eq.(6).

 The DBD isotopes of  $^{82}$Se,  $^{100}$Mo  and $^{136}$Xe are  very useful isotopes because of the small  $m'_0 \approx$ 40 meV due to the large phase space  $G^{0\nu}$ and the large  $Q$ value and multi-ton scale enriched isotopes with $N \approx$10 and $\eta \approx $90$\%$ are available  by centrifugal separation.
 
 Cryogenic $^{82}$Se- and $^{100}$Mo- bolometers with both thermal and scintillation signals are promising detectors with  high energy resolution and low BG rate. See refs. \cite{art17,par16,pod17,don20}. $^{116}$Cd with $m'_0\approx$40 mev is also interesting. See refs. \cite{zub01,dan16}. 
 
 $^{136}$Xe can be easily  enriched to get a 10 t scale isotope-mass and thus is possible to access  the IH and NH masses. Various kinds of $^{136}$Xe detectors are under progress. See refs. \cite{gan16,alb18,yan17,ant19,gra20,gom20,mar18}.
 
 $^{76}$Ge with $Q$=2.039 MeV has the unit mass of $m'_0$=80 meV, a factor 2 larger than the others given above. Ton scale enriched DBD isotopes are possible by centrifugal separation, and  the energy resolution of 10$^{-3}$ in FWHM is very good. Accordingly the Ge  BG rate is almost an order of magnitude  smaller than others to get the neutrino mass sensitivity comparable with other detectors. See refs. \cite{ago18,aal18,alv19,abg17,ker20}.
 
 The natural abundance of $^{130}$Te is 34$\%$, and thus multi-ton scale natural Te isotopes may be used although the mass sensitivity $m_m$ gets larger by a factor 1.6 than the mass sensitivity with enriched $^{130}$Te-isotope detectors. 
  See ref. \cite{don20,ada19}
 
$^{150}$Nd has a large phase space factor  because of the large $Q$ value of 3.4 MeV and the large $Z$ number of 60. Thus the $m'_0$=18 meV is  a half of the others \cite{eji19,eji20}. The natural abundance, however, is only 5.6 $\%$ and the enrichment is not easy. See ref.\cite{and16}.
 
 Tracking detectors have been used to study DBDs on $^{100}$Mo, $^{116}$Cd, $^{82}$Se, and many other DBD nuclei as discussed in review articles and references therein \cite{eji05,avi08,ver12,ver16}. They measure individual two $\beta $-rays to identify the DBD signals and to reject  single-$\beta$ background signals. The measured  energy and angular correlations  are used to differentiate the DBD processes due to the left-handed and right-handed weak currents. See refs. \cite{eji00a, eji08,wat17}.  
  
Finally it should be remarked that the NME is one of key ingredients for the DBD mass sensitivity, and thus selection of DBD isotopes with a large NME is very crucial for getting high-sensitivity DBD experiments \cite{eji00,eji19}. It is however very hard to evaluate accurately the DBD NMEs. So various experiments are under progress to provide nuclear parameters, the effective axial-vector weak couplings and nuclear structures to help theoretical evaluations of the neutrinoless DBD NMEs.  These are discussed in recent articles \cite{eji20,eji21}.\\

{\bf Acknowledgement}

The author would like to express many thanks to Prof. Haris Kosmas for valuable discussions and encouragements and
best wishes for his future activities on the occasion of his 70$^{th}$ birthday. \\

\end{document}